\documentclass{article}
\usepackage{spconf,amsmath,graphicx}
\usepackage{subcaption}
\usepackage{amssymb}
\usepackage{bbm}
\usepackage{algorithm}
\usepackage{algpseudocode}
\usepackage{url}
\usepackage{hyperref}
\usepackage{booktabs}
\usepackage{enumitem}

\usepackage{caption}
\title{Profile-Aware Trustworthy Recipe Generation with Planner-Critic Agentic Remediation}
\name{Shanhong Liu$^*$, Pai Chet Ng$^\ddagger$, Konstantinos N. Plataniotis$^\dagger$ \vspace{-0.5cm}}
\address{ 
\normalsize$^*$Information Systems Technology and Design, Singapore University of Technology and Design, Singapore \\
\normalsize$^\ddagger$Infocomm Technology Cluster, Singapore Institute of Technology, Singapore \\
\normalsize$^\dagger$The Edward S. Rogers Sr. Department of Electrical \& Computer Engineering, University of Toronto, Canada \\
}
\begin{document}
%
\maketitle
\begin{abstract}
Recipe generation from food images has practical value for intelligent cooking assistance, but traditional one-pass generation often overlooks user-specific safety requirements such as allergies, dietary restrictions, and preparation constraints. We propose \textbf{PCAR}, a \textbf{P}lanner-\textbf{C}ritic \textbf{A}gentic \textbf{R}emediation framework for trustworthy recipe generation. PCAR separates recipe planning from safety verification: a Planner Agent extracts ingredients and generates recipe drafts conditioned on the user profile, while a Safety Critic Agent audits each draft and provides structured feedback for remediation when violations are detected. This remediation loop enables unsafe recipes to be revised rather than directly returned or discarded. We evaluate PCAR on real food images with 100 benchmark user profiles across four backbone models, including proprietary and locally served open-source models. Results show that PCAR achieves strong safety and generation performance with capable backbone models, while preserving practical recipe quality. 
\end{abstract}
\section{Introduction}
Food image understanding~\cite{8998172, 10019590, 9594708} has attracted increasing attention due to its potential applications in nutrition analysis and intelligent cooking assistance. Prior studies have explored food recognition and ingredient identification from visual inputs, including ingredient-aware recommendation systems~\cite{10.1145/3542954.3542983, gao2019hierarchical}, dietary tracking~\cite{ming2018food, 9082900}, recipe retrieval~\cite{10.1145/2964284.2964315, 10494044}, multi-task ingredient recognition~\cite{9305995, 10471331}, recipe generation~\cite{Papadopoulos_2022_CVPR,9793670} and region-wise ingredient modeling~\cite{9846887, Wu_2024_CVPR}. These works have significantly advanced the ability to recognize food categories and infer ingredients from images. More recently, large language models have shown promise for food-related applications, including food entity extraction, fine-grained food recognition, and food-oriented reasoning and generation~\cite{10555071,10.1145/3607828.3617797}. 
However, existing works are still limited when deployed as personalized cooking assistants, as they pays less attention to whether the generated recipe is safe for a specific user.

\begin{figure}[t]
    \centering
    \includegraphics[width=\columnwidth]{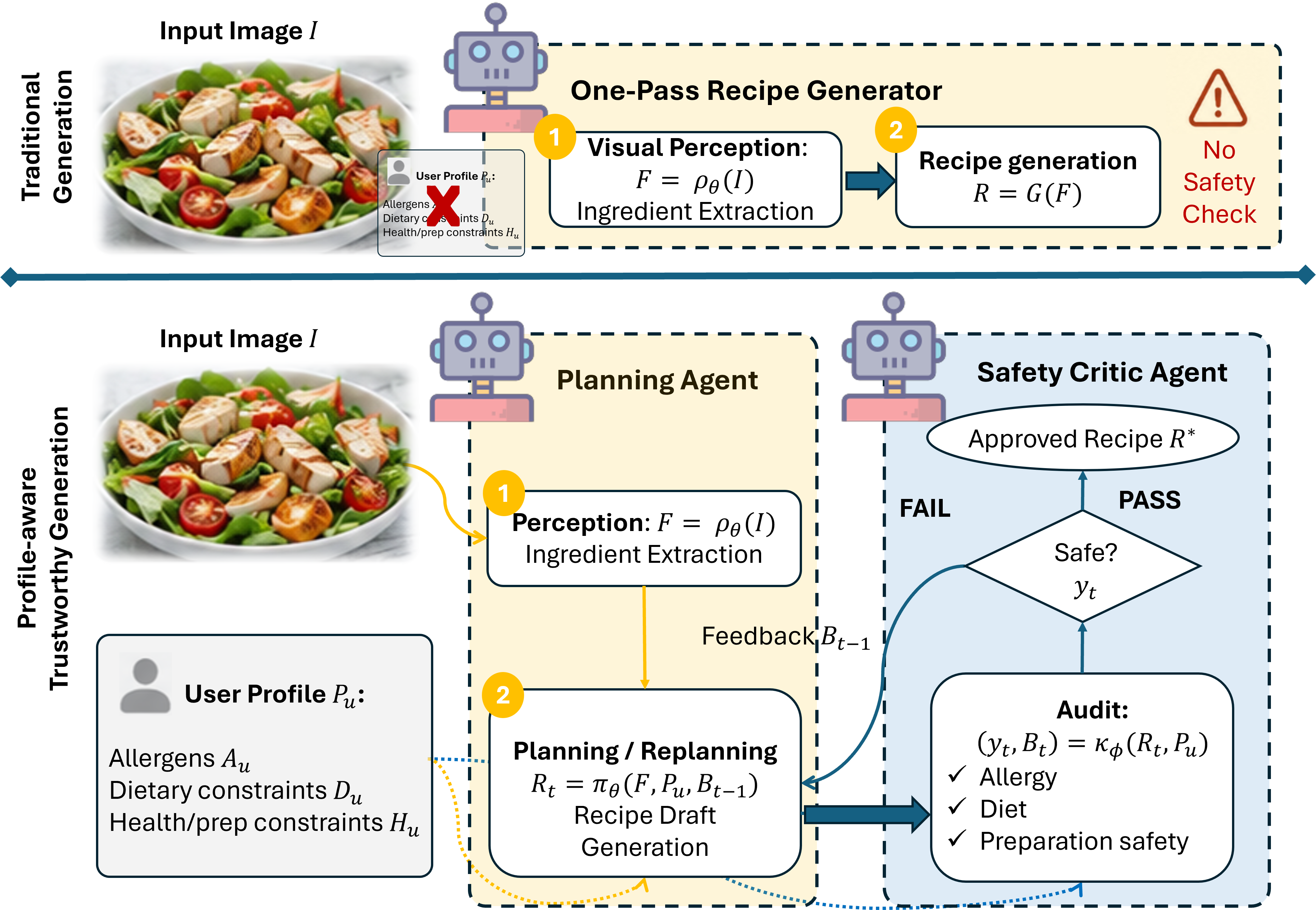}
    \caption{
    Traditional one-pass recipe generation maps a food image directly to a recipe without enforcing user-profile constraints or performing safety verification. In contrast, we propose a profile-aware planner--critic remediation loop, where the Planner Agent generates and revises recipe drafts, while the Safety Critic Agent audits each draft against allergy, dietary, and preparation-safety constraints before approval.
    }
    \label{fig:pcar_intro_overview}
    \vspace{-0.35cm}
\end{figure}

In practice, recipe generation is not merely a generative task; it is also a safety-sensitive decision process. A generated recipe may include allergens, violate dietary restrictions, or omit critical preparation-safety steps. For example, a visually plausible recipe may still be unsafe for a user with a nut allergy, incompatible with vegan or halal requirements, or inappropriate for vulnerable users such as infants or pregnant individuals. As illustrated in Fig.~\ref{fig:pcar_intro_overview}, a traditional one-pass generation maps a food image directly to a recipe, and does not explicitly enforce user-profile constraints, or verify safety, or provide a mechanism for revising unsafe outputs. Simply adding user information to the prompt is also insufficient, as the model may still overlook constraints or fail to detect hidden safety risks. Similarly, a post-hoc safety filter can reject unsafe recipes, but it does not recover a useful safe alternative.

To address these limitations, we propose \textbf{PCAR}, a \textbf{P}lanner-\textbf{C}ritic \textbf{A}gentic \textbf{R}emediation framework for profile-aware trustworthy recipe generation, casting recipe generation as a constrained generation process that balances practical recipe utility with profile-aware safety satisfaction. Specifically, PCAR decomposes recipe generation into two interacting agents: (1) A Planner Agent performs visual ingredient understanding and generates recipe drafts conditioned on the food image and user profile; and (2) A Safety Critic Agent independently audits each draft against user-specific allergens, dietary constraints, and preparation-safety requirements. When a draft is unsafe, the critic returns structured feedback to the planner, enabling the recipe to be revised rather than simply rejected. The planner--critic loop continues until the recipe is approved or a bounded remediation limit is reached.  
Our contributions are summarized as follows:
\begin{itemize}[noitemsep, topsep=0pt]
    \item We formulate recipe generation as a constrained generation objective, where the generated recipe should remain useful and image-grounded while satisfying user-specific allergy, dietary, and preparation-safety constraints.
    \item Our PCAR enforces an agentic remediation mechanism in which unsafe recipe drafts are not merely rejected, but revised using structured feedback from an independent Safety Critic Agent.
    \item We evaluate PCAR across four backbone models, including proprietary and locally served open-source models, using real food images and 100 benchmark user profiles spanning baseline constraints, complex combinations, derivative risks, and high-risk scenarios.
\end{itemize}
\section{Proposed Method}
\label{sec:method}

\subsection{Problem Formulation}
\label{subsec:problem_formulation}
A traditional one-pass recipe generation system can be abstracted as
\begin{equation}
    R = G(I),
    \label{eq:one_pass_generation}
\end{equation}
where a single model maps the food image $I$ directly to a recipe $R$. Such a formulation entangles visual understanding and recipe generation, making the output vulnerable to user constraints, hallucinated ingredients, or unsafe preparation steps. We therefore formulate profile-aware recipe generation as a constrained generation objective: the output recipe $R$ should be practically useful while satisfying user-specific safety constraints,
\begin{equation}
    R^\star \in \mathcal{R}_{\mathrm{safe}}(P_u)
    \quad \text{and} \quad
    U(R^\star; I, P_u) \text{ is high},
    \label{eq:constrained_generation_objective}
\end{equation}
where $U(R; I, P_u)$ denotes the utility of the recipe and $\mathcal{R}_{\mathrm{safe}}(P_u)$ denotes the set of recipes that are safe for user $u$. 

Let user profile be $P_u = (A_u, D_u, H_u)$, 
where $A_u$ is the set of user-specific allergens, $D_u$ denotes dietary restrictions, and $H_u$ represents additional health or preparation-safety constraints. 
We define the safe recipe space for user $u$ as
\begin{equation}
    \mathcal{R}_{\mathrm{safe}}(P_u)
    =
    \{R \in \mathcal{R} \mid S(R, P_u)=1\},
    \label{eq:safe_recipe_space}
\end{equation}
where $S(R,P_u)\in\{0,1\}$ indicates whether $R$ satisfies all profile-aware and preparation-aware safety requirements. We decompose $S(R,P_u)$ as
\begin{equation}
    S(R, P_u) =
    S_{\mathrm{allergy}}(R, A_u)
    \land
    S_{\mathrm{diet}}(R, D_u)
    \land
    S_{\mathrm{prep}}(R, H_u),
    \label{eq:safety_decomposition}
\end{equation}
where the three terms ($S_{\mathrm{allergy}}$, $S_{\mathrm{diet}}$, and $S_{\mathrm{prep}}$) check allergen avoidance, dietary compliance, and safe preparation procedures, respectively. The utility term $U(R; I, P_u)$ captures non-safety qualities such as image grounding, culinary coherence, instruction completeness, and personalization.

\subsection{Planner--Critic Agentic Remediation (PCAR)}
\label{subsec:pcar}
To address the constrained generation objective in Eq.~\eqref{eq:constrained_generation_objective}, 
our PCAR casts recipe generation as an iterative constrained generation process by decoupling recipe generation from safety verification, PCAR . 
Specifically, PCAR separates these responsibilities into two interacting agents: 
(1) a Planner Agent to propose and revise recipe drafts and 
(2) a Safety Agent to audit each draft against the user profile and provides structured feedback when violations are detected.


\noindent \textbf{Planner Agent.}
The Planner Agent generates image-grounded and profile-aware recipe drafts. Let $\pi_{\theta}$ denote its planning policy and $\rho_{\theta}$ its perception module. Given an image $I$, the planner first extracts perceived ingredients $F=\rho_{\theta}(I)$, then generates a draft $R_t$ conditioned on $F$, the user profile $P_u$, and any previous safety feedback $B_{t-1}$.
Mathematically, the recipe draft $R_t$ at time step $t$ is described as
\begin{equation}
    R_t =
    \begin{cases}
    \pi_{\theta}(F, P_u, \emptyset), & t = 0, \\
    \pi_{\theta}(F, P_u, B_{t-1}), & t \geq 1,
    \end{cases}
    \label{eq:planner_policy}
\end{equation}
where the initial step produces a profile-aware recipe from the perceived ingredients, while later steps perform feedback-conditioned replanning. During remediation, the planner uses $B_{t-1}$ to revise unsafe ingredients, substitutions, or preparation steps without discarding the recipe's image grounding and practical utility.

\noindent \textbf{Safety Agent.}
Given a recipe draft \(R_t\) and user profile \(P_u\), the Safety Critic Agent audits whether the draft satisfies the safety constraints in Eq.~\eqref{eq:safety_decomposition}. Let \(\kappa_{\phi}\) denote its auditing policy. The critic outputs a binary safety decision \(y_t\) and feedback \(B_t\):
\begin{equation}
    (y_t, B_t) = \kappa_{\phi}(R_t, P_u).
    \label{eq:safety_critic_policy}
\end{equation}
The decision is based on the violation set $\mathcal{V}_t=
    \mathcal{V}_{\mathrm{allergy}}(R_t,A_u) \\
    \cup
    \mathcal{V}_{\mathrm{diet}}(R_t,D_u)
    \cup
    \mathcal{V}_{\mathrm{prep}}(R_t,H_u)$,  
which captures allergen $\mathcal{V}_{\mathrm{allergy}}$, dietary $\mathcal{V}_{\mathrm{diet}}$, and preparation-safety $\mathcal{V}_{\mathrm{prep}}$ violations. 
This binary safety decision can be defined as follows:
\begin{equation}
    y_t =
    \begin{cases}
    1, & \text{if } |\mathcal{V}_t| = 0, \\
    0, & \text{if } |\mathcal{V}_t| > 0.
    \end{cases}
    \label{eq:safety_decision}
\end{equation}
When $y_t=1$, the recipe draft $R_t \in \mathcal{R}_{\mathrm{safe}}(P_u)$ is considered safe. 
When $y_t=0$, the recipe is unsafe as it violates at least one safety condition.
For unsafe drafts, the Safety Agent further produces a structured feedback, 
\begin{equation}
    B_t = (v_t, e_t),
    \label{eq:feedback}
\end{equation}
where \(v_t\) denotes the violation type and \(e_t\) explains the safety issue, such as identifying an ingredient that conflicts with the user's allergen set.

\noindent \textbf{Bounded Planner--Safety Remediation Loop.}
The interaction between the Planner Agent and the Safety Agent forms a closed-loop remediation process
that operationalizes the constrained generation objective in Eq.~\eqref{eq:constrained_generation_objective}. The planner aims to produce recipes with high utility $U(R;I,P_u)$, while  Safety Agent enforces the safety condition $S(R,P_u)=1$ by detecting violations and generating corrective feedback. Unsafe drafts are therefore not accepted directly, but iteratively revised toward \(\mathcal{R}_{\mathrm{safe}}(P_u)\). 
To prevent unbounded agentic execution, PCAR imposes a maximum remediation budget $T$, with the termination policy is defined as
\begin{equation}
    \tau(R_t, y_t, t)=
    \begin{cases}
    \mathrm{approve}(R_t), & y_t = 1, \\
    \mathrm{remediate}(B_t), & y_t = 0 \land t < T, \\
    \mathrm{fail\mbox{-}safe}, & y_t = 0 \land t = T.
    \end{cases}
    \label{eq:termination_policy}
\end{equation}
This bounded policy  ensures that unsafe recipes are not returned to the user, while also preventing the system from entering an indefinite correction loop. 
The complete procedure is summarized in Algorithm~\ref{alg:pcar}.


\begin{algorithm}[t]
\caption{PCAR: Profile-Aware Planner--Critic Agentic Remediation}
\label{alg:pcar}
\begin{algorithmic}[1]
\Require Food image $I$, user profile $P_u$, maximum remediation budget $T$
\Ensure Approved recipe $R^\star$ or fail-safe termination
\State Extract perceived ingredients $F \leftarrow \rho_{\theta}(I)$
\State Initialize safety feedback $B_{-1} \leftarrow \emptyset$
\For{$t = 0, \ldots, T$}
    \State Generate recipe draft $R_t \leftarrow \pi_{\theta}(F, P_u, B_{t-1})$
    \State Audit recipe $(y_t, B_t) \leftarrow \kappa_{\phi}(R_t, P_u)$
    \If{$y_t = 1$}
        \State \Return $R^\star \leftarrow R_t$
    \EndIf
\EndFor
\State \Return fail-safe termination
\end{algorithmic}
\end{algorithm}

\section{Experiments and Evaluation}
\label{sec:experiments}

\subsection{Experimental Setup}
\label{subsec:experimental_setup}
We construct an experimental setting that combines real food images with a controlled suite of user profiles. The food images provide realistic visual inputs for ingredient perception and recipe generation, while the user profiles introduce profile-aware safety constraints for evaluating our PCAR framework.

\noindent \textbf{Food image dataset.}
We used the Epicurious recipe dataset as the food-image source for our experiments\footnote{https://www.kaggle.com/datasets/hugodarwood/epirecipes}. 
The dataset contains over 20K recipes with metadata, and we use the version with food images and cleaned ingredient lists. Each image serves as the visual input to PCAR, while the cleaned ingredient list is used only as the reference for evaluating ingredient perception. We do not use the original recipes as generation targets, since our goal is not recipe reconstruction but profile-aware recipe generation that is visually grounded and satisfies user-specific safety constraints.

\noindent \textbf{User Profiles.}
To evaluate profile-aware safety, we construct a benchmark suite of 100 user profiles, each containing an allergy set $A_u$, a dietary regimen $D_u$, and a natural-language constraint description. 
The benchmark suite is organized into four cohorts, with 25 profiles each:
\begin{itemize}[noitemsep, topsep=0pt]
    \item \textbf{Baseline Control:} isolated allergies or standard diets, such as peanuts, shellfish, vegetarian, vegan, and halal.
    \item \textbf{Complex Combinations:} multiple simultaneous constraints, such as vegan with nut allergies or halal with wheat restriction.
    \item \textbf{Derivative Risks:} hidden or indirect allergens, such as fish sauce for seafood allergy.
    \item \textbf{High-Risk Scenarios:} vulnerable or preparation-sensitive profiles, such as pregnancy-safe, infant-safe, low-sodium, elderly-soft, and toxin-test constraints.
\end{itemize}
The first two cohorts test explicit constraint following, while the latter two evaluate whether PCAR can identify less obvious safety risks and provide corrective feedback through the planner--critic remediation loop.

\begin{table*}[t]
\centering
\small
\caption{Overall performance of PCAR across backbone models. Results are averaged over 300 cases per model.  
}
\label{tab:overall_performance}
\resizebox{\linewidth}{!}{%
\begin{tabular}{l l c c c c c c c c}
\toprule
\textbf{Model} 
& \textbf{Type}
& \textbf{Crash} $\downarrow$
& \textbf{SGSR} $\uparrow$
& \textbf{ULR} $\downarrow$
& \textbf{RASS} $\uparrow$
& \textbf{MRL} $\downarrow$
& \textbf{Latency (s)} $\downarrow$
& \textbf{Sem. F1} $\uparrow$
& \textbf{Quality} $\uparrow$ \\
\midrule
GPT-4o 
& Proprietary 
& 0.00 
& 98.00 
& 2.00 
& 0.9783 
& 1.01 
& 6.07 
& 0.0487 
& 4.2982 \\

Gemini-2.5-Flash 
& Proprietary 
& 14.67 
& 74.00 
& 11.33 
& 0.7183 
& 0.91 
& 17.01 
& 0.0456 
& 4.5357 \\

Gemma3-4B-IT 
& Local open-source 
& 34.00 
& 39.67 
& 7.00 
& 0.2817 
& 1.36 
& 33.10 
& 0.0180 
& 4.6667$^\dagger$ \\

Qwen2.5-VL-3B 
& Local open-source 
& 47.00 
& 10.00 
& 0.00 
& 0.0606 
& 1.49 
& 11.43 
& 0.0120 
& 3.4000$^\dagger$ \\
\bottomrule
\end{tabular}
}
\vspace{-0.35cm}

\end{table*}
\subsection{Evaluation protocol}
\label{subsec:evaluation_protocol}
We evaluate PCAR across four backbone models. For each model, we run three trials over 100 user profiles, yielding 300 test cases. Each case pairs one randomly sampled Epicurious food image with one user profile to assess visual grounding, profile compatibility, safety, and practical recipe quality.

\noindent \textbf{Backbone models.}
Two proprietary API-based models (\texttt{gpt-4o}~\footnote{https://openai.com/index/hello-gpt-4o/} and \texttt{gemini-2.5-flash}~\footnote{https://ai.google.dev/gemini-api/docs/models/gemini-2.5-flash}) and two locally served open-source models (\texttt{qwen2\_5\_vl\_3b}~\footnote{https://huggingface.co/Qwen/Qwen2.5-VL-3B-Instruct} and \texttt{gemma3-4b-it}~\footnote{https://huggingface.co/google/gemma-3-4b-it}) are deployed as backbone  to compare PCAR across both high-capacity cloud models and lightweight local models. We use relatively small local models because the intended deployment scenario of trustworthy recipe generation may require consumer-grade or edge-accessible hardware rather than large-scale GPU infrastructure. Accordingly, all local-model experiments are conducted on a workstation with an NVIDIA RTX 4090 GPU.

\noindent \textbf{Evaluation Metrics}
Our goal is to assess ingredient perception, recipe generation that satisfy user-specific constraints along five dimensions::
\begin{itemize}[noitemsep, topsep=0pt]
    \item \textbf{Perception quality.}
    We evaluate the Planner Agent's ingredient perception by comparing the predicted ingredient set \(F_i\) with the cleaned reference set \(F_i^{\mathrm{gt}}\). Since exact matching can penalize equivalent surface forms, we use semantic ingredient F1 as the primary metric. Semantic true positives are computed by thresholded one-to-one matching:
    \begin{equation}
    \begin{aligned}
        \mathrm{TP}_i^{\mathrm{sem}}
        =
        \max_{\mathbf{X}}
        \sum_{p=1}^{|F_i|}
        \sum_{q=1}^{|F_i^{\mathrm{gt}}|}
        X_{pq}\mathbf{1}[s_{pq}\geq\delta], \\
        \text{s.t. } \quad
        \sum_q X_{pq}\leq 1,\;
        \sum_p X_{pq}\leq 1,
    \end{aligned}
    \label{eq:semantic_tp}
    \end{equation}
    where $s_{pq}
        =
        \cos \bigl(z(f_{i,p}), z(f^{\mathrm{gt}}_{i,q})\bigr.$ 
    is the embedding-based similarity, \(X_{pq}\in\{0,1\}\) indicates a matched pair, and \(\delta\) is the similarity threshold. We greedily match candidate pairs satisfying \(s_{pq}\geq\delta\), and use \(\mathrm{TP}_i^{\mathrm{sem}}\) to compute precision, recall, and F1.

    \item \textbf{Safe generation success and leakage.}
    A recipe is counted as a successful generation if it is approved by the Safety Critic Agent and passes the lexical allergen leakage check.
    The safe generation success rate is
    \begin{equation}
        \mathrm{SGSR}
        =
        \frac{1}{N}
        \sum_{i=1}^{N}  \mathbf{1}
        [
        y_i^\star = 1
        \land
        R_i^\star \neq \emptyset
        \land
        \ell_i = 0
        ],
    \end{equation}
    where $N$ is the number of evaluated cases, $y_i^\star$ is the final safety decision, $R_i^\star$ is the final recipe, and $\ell_i$ indicates whether a forbidden allergen is detected. 
    We also compute the unsafe leakage rate as
    \begin{equation}
        \mathrm{ULR}
        =
        \frac{1}{N}
        \sum_{i=1}^{N}
        \mathbf{1}[y_i^\star = 1 \land \ell_i = 1].
    \end{equation}


    \item \textbf{Remediation efficiency.}
    To measure remediation efficiency, we record the number of critic evaluations \(c_i\) for each case and compute the mean remediation count:
    \begin{equation}
        \mathrm{MRL}
        =
        \frac{1}{N}
        \sum_{i=1}^{N} c_i.
    \end{equation}
    We also compute a remediation-adjusted safe success score, which rewards safe generation while penalizing excessive remediation:
    \begin{equation}
        \mathrm{RASS}_i =
        \begin{cases}
        \frac{1}{\max(c_i,1)}, & \text{if } g_i = 1, \\
        0, & \text{otherwise}.
        \end{cases}
    \end{equation}
    The final RASS is averaged over all $\mathrm{RASS}_i$.

    \item \textbf{Recipe quality.}
    For successful approved recipes, we use an independent LLM-based judge to assign 1--5 scores for: \textit{culinary coherence}, \textit{instruction specificity}, \textit{ingredient grounding}, \textit{constraint compatibility}, \textit{practical feasibility}, and \textit{overall quality}. Failed, crashed, or unsafe cases are excluded from the assessment because they do not produce a valid final recipe. 

    \item \textbf{Efficiency and latency.}
    We measure end-to-end latency for each profile-image case, covering perception, planning, safety criticism, and remediation. Together with mean remediation loops, latency characterizes the safety--efficiency trade-off across backbone models.
\end{itemize}

\begin{table*}[t]
\centering
\small
\caption{
Cohort-level robustness of PCAR across user-profile cohorts. Each cell reports \textbf{SGSR / ULR} in percentage.
}
\label{tab:cohort_robustness}
\begin{tabular}{l c c c c}
\toprule
\textbf{Model} 
& \textbf{Baseline Control}
& \textbf{Complex Combinations}
& \textbf{Derivative Risks}
& \textbf{High-Risk Scenarios} \\
\midrule
GPT-4o 
& 100.00 / 0.00 
& 93.33 / 6.67 
& 98.67 / 1.33 
& 100.00 / 0.00 \\

Gemini-2.5-Flash 
& 81.33 / 5.33 
& 65.33 / 20.00 
& 69.33 / 20.00 
& 80.00 / 0.00 \\

Gemma3-4B-IT 
& 40.00 / 5.33 
& 38.67 / 8.00 
& 48.00 / 14.67 
& 32.00 / 0.00 \\

Qwen2.5-VL-3B 
& 17.33 / 0.00 
& 5.33 / 0.00 
& 16.00 / 0.00 
& 1.33 / 0.00 \\
\bottomrule
\end{tabular}
\end{table*}

\begin{table*}[t]
\centering
\small
\caption{
Recipe quality of successful approved generations. Scores are assigned by an independent LLM-based judge on a 1--5 scale. 
}
\label{tab:recipe_quality}
\begin{tabular}{l c c c c c c c}
\toprule
\textbf{Model}
& \textbf{Judged}
& \textbf{Coherence}
& \textbf{Specificity}
& \textbf{Grounding}
& \textbf{Compatibility}
& \textbf{Feasibility}
& \textbf{Overall} \\
\midrule
GPT-4o
& 57
& 4.9825
& 4.5088
& 4.2982
& 4.5263
& 5.0000
& 4.2982 \\

Gemini-2.5-Flash
& 84
& 4.7500
& 4.4762
& 4.4524
& 4.8929
& 4.7619
& 4.5357 \\

Gemma3-4B-IT
& 3
& 4.6667
& 4.3333
& 4.6667
& 5.0000
& 5.0000
& 4.6667 \\

Qwen2.5-VL-3B
& 5
& 4.0000
& 2.8000
& 4.2000
& 4.6000
& 4.8000
& 3.4000 \\
\bottomrule
\end{tabular}

\vspace{-0.5em}
\end{table*}
\begin{figure}[t]
    \centering
    \includegraphics[width=.9\linewidth]{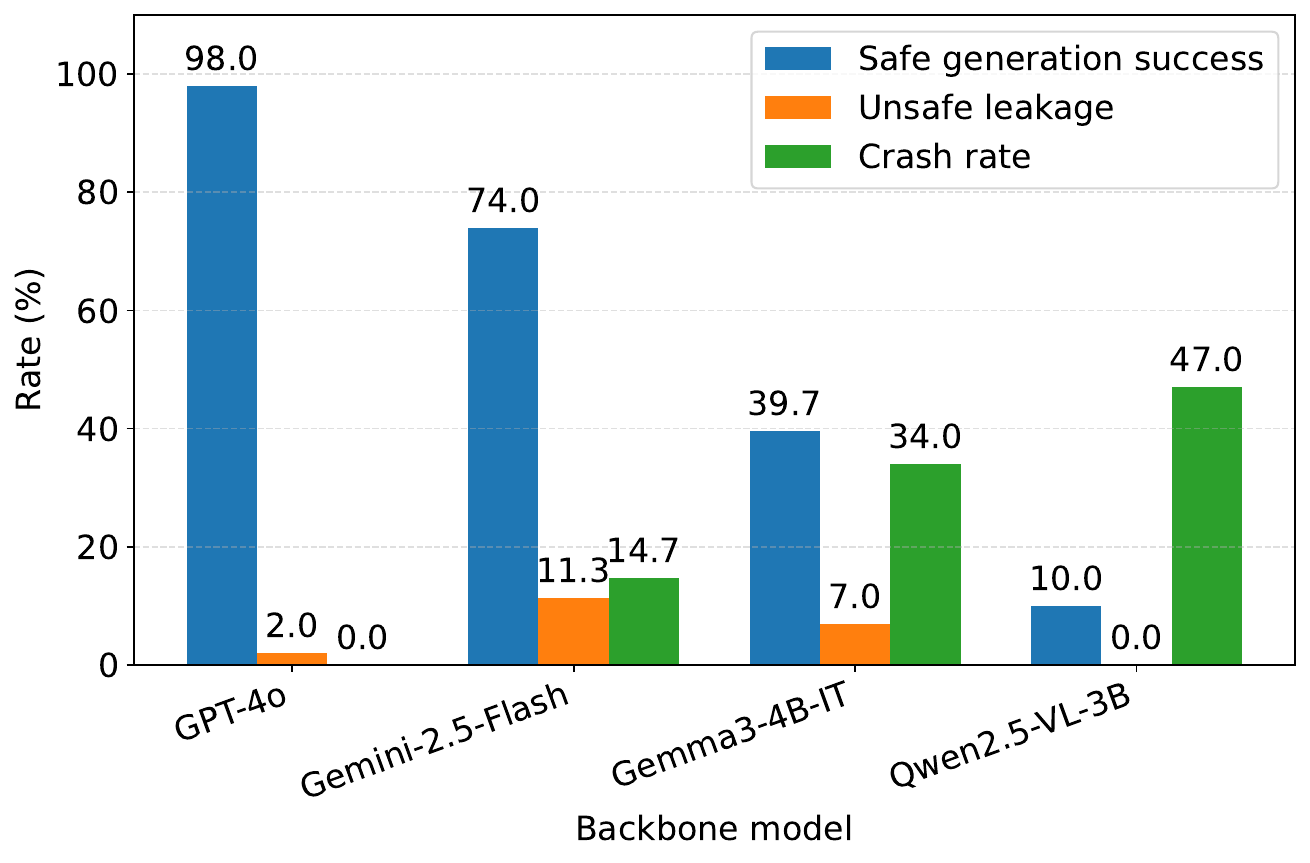}
    \caption{
    Safety performance of PCAR across backbone models.}
    \label{fig:safety_grouped_bar}
    \vspace{-0.35cm}
\end{figure}
\subsection{Results and Discussion}
We organize the results around four research questions:
\textbf{RQ1:} Can PCAR generate safe profile-compatible recipes?
\textbf{RQ2:} How does backbone choice affect safety and efficiency?
\textbf{RQ3:} How robust is PCAR across user-profile cohorts?
\textbf{RQ4:} Do approved recipes preserve practical quality?

\noindent \textbf{RQ1: Overall safety and generation performance.}
As shown in Table~\ref{tab:overall_performance} and Fig.~\ref{fig:safety_grouped_bar},  GPT-4o achieves the strongest result, with $98\%$ safe generation success, only $2\%$ unsafe leakage, and no execution crashes. This indicates that PCAR can reliably generate profile-compatible recipes when supported by a strong instruction-following backbone. Gemini-2.5-Flash also demonstrates reasonable safe generation capability, achieving $74\%$ SGSR, but its higher unsafe leakage ($11.33\%$) and crash rate ($14.67\%$) suggest weaker stability under the same planner--critic workflow. In contrast, the locally served models show lower robustness: Gemma3-4B-IT reaches $39.67\%$ SGSR, while Qwen2.5-VL-3B reaches only $10\%$, both with substantially higher crash rates. These results suggest that PCAR is a viable but its reliability depends strongly on the backbone model's structured-output stability, safety reasoning, and ability to follow remediation feedback.

\begin{figure}[t]
    \centering
    \includegraphics[width=0.85\linewidth]{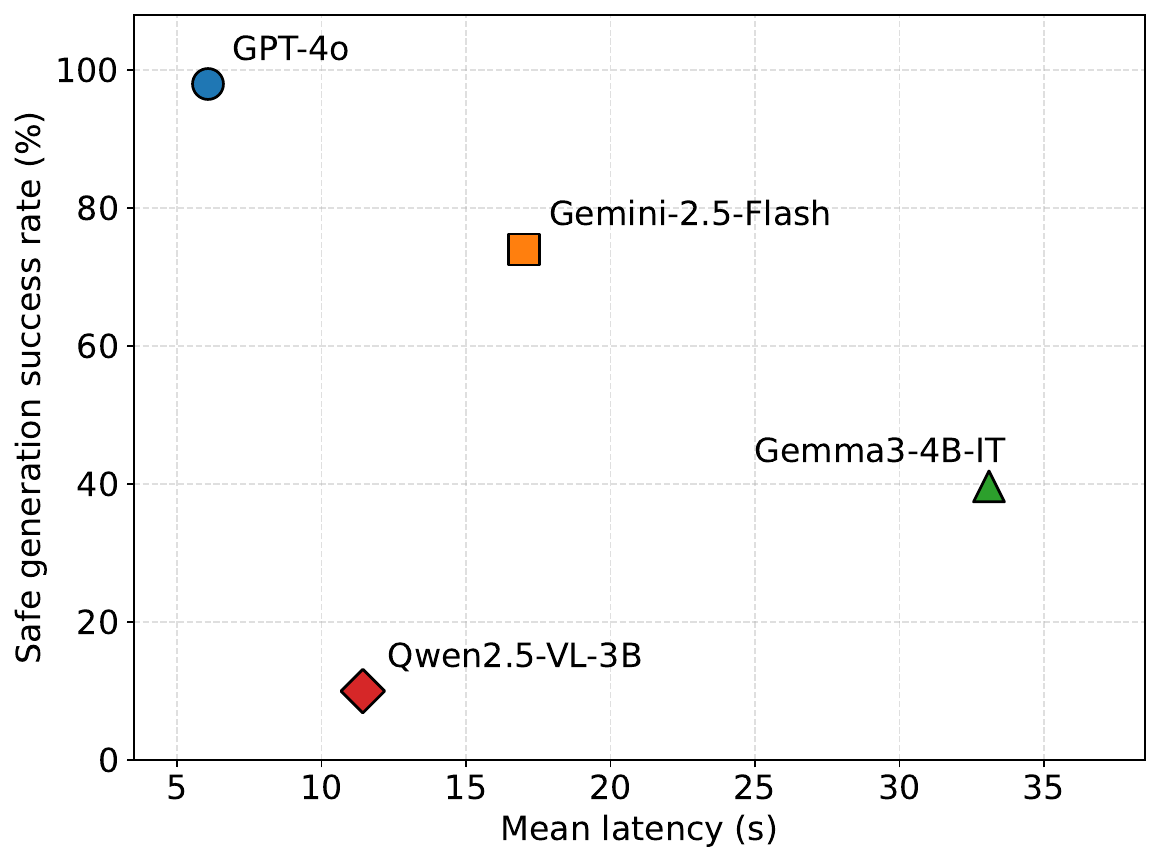}
    \caption{
    Safety--efficiency trade-off of PCAR across backbones. 
    }
    \label{fig:safety_efficiency_tradeoff}
\end{figure}

\noindent \textbf{RQ2: Safety--efficiency trade-off.}
Fig.~\ref{fig:safety_efficiency_tradeoff} shows the safety--efficiency trade-off. GPT-4o achieves both the highest safe generation success and the lowest latency, suggesting that stronger instruction following and structured generation reduce remediation overhead.
Among local models, Qwen2.5-VL-3B is faster than Gemma3-4B-IT but much less safe, while Gemma3-4B-IT remains slower and less reliable than proprietary models. This suggests that local deployment is feasible, but small models still struggle with safety and stability in multi-step remediation.

\noindent \textbf{RQ3: Profile-cohort robustness.}
Table~\ref{tab:cohort_robustness} reports cohort-level robustness across the four user-profile cohorts. GPT-4o remains consistently strong across all cohorts, achieving $100\%$ SGSR with no unsafe leakage on both Baseline Control and High-Risk Scenarios, and maintaining high success on Complex Combinations and Derivative Risks. This indicates that PCAR can handle both explicit user constraints and more challenging safety profiles when the backbone model is sufficiently capable. 
The local models show much lower cohort-level robustness, especially Qwen2.5-VL-3B on Complex Combinations and High-Risk Scenarios. These results confirm that profile-aware trustworthy recipe generation is not only a matter of adding user profiles to prompts; the system must also reliably interpret constraint interactions and convert safety critiques into valid revisions.

\noindent \textbf{RQ4: Recipe quality analysis.}
According to the results shown in Table~\ref{tab:recipe_quality}, Gemini-2.5-Flash obtains the highest overall quality score ($4.5357$), while GPT-4o achieves strong culinary coherence ($4.9825$) and perfect practical feasibility ($5$), validates that recipes approved by PCAR are generally coherent, feasible, and compatible with user constraints. However, recipe quality must be interpreted together with safe generation success. Although Gemma3-4B-IT receives a high average quality score, only three successful recipes are judged, making the estimate unstable. Similarly, Qwen2.5-VL-3B has only five judged successful recipes and a lower overall quality score. Therefore, the recipe-quality results mainly show that when PCAR successfully produces an approved recipe, the output can be practically useful.

\section{Conclusion}
This work demonstrates that trustworthy recipe generation requires more than producing visually plausible recipes: it must also account for user-specific dietary, allergy, and preparation-safety constraints. 
Our results indicate that PCAR can achieve high safe generation success while preserving practical recipe quality, but they also reveal important limitations: performance remains highly dependent on the underlying model, lightweight local models show weaker stability. 
Future work will focus on strengthening local-model reliability, incorporating richer nutrition and medical constraint modeling, and  improving ingredient perceptions.


\bibliographystyle{IEEEbib}
\bibliography{strings,refs}

@String{Computing = "Computing" }

@String{Computer = "{IEEE} Computer" }

@String{Springer = "Springer-Verlag" }

@ArtifactSoftware{R,
    title = {R: A Language and Environment for Statistical Computing},
    author = {{R Core Team}},
    organization = {R Foundation for Statistical Computing},
    address = {Vienna, Austria},
    year = {2019},
    url = {https://www.R-project.org/},
}

@inproceedings{10.1145/3542954.3542983,
author = {Morol, Md. Kishor and Rokon, Md. Shafaat Jamil and Hasan, Ishra Binte and Saif, A. M. and Khan, Rafid Hussain and Das, Shuvra Smaran},
title = {Food Recipe Recommendation Based on Ingredients Detection Using Deep Learning},
year = {2022},
isbn = {9781450397346},
publisher = {Association for Computing Machinery},
address = {New York, NY, USA},
url = {https://doi.org/10.1145/3542954.3542983},
doi = {10.1145/3542954.3542983},
booktitle = {Proceedings of the 2nd International Conference on Computing Advancements},
pages = {191–198},
numpages = {8},
location = {Dhaka, Bangladesh},
series = {ICCA '22}
}

@inproceedings{10.1145/2964284.2964315,
author = {Chen, Jingjing and Ngo, Chong-wah},
title = {Deep-based Ingredient Recognition for Cooking Recipe Retrieval},
year = {2016},
isbn = {9781450336031},
publisher = {Association for Computing Machinery},
address = {New York, NY, USA},
url = {https://doi.org/10.1145/2964284.2964315},
doi = {10.1145/2964284.2964315},
booktitle = {Proceedings of the 24th ACM International Conference on Multimedia},
pages = {32–41},
numpages = {10},
location = {Amsterdam, The Netherlands},
series = {MM '16}
}

@ARTICLE{9305995,
  author={Chen, Jingjing and Zhu, Bin and Ngo, Chong-Wah and Chua, Tat-Seng and Jiang, Yu-Gang},
  journal={IEEE Transactions on Image Processing}, 
  title={A Study of Multi-Task and Region-Wise Deep Learning for Food Ingredient Recognition}, 
  year={2021},
  volume={30},
  number={},
  pages={1514-1526},
  doi={10.1109/TIP.2020.3045639},
  ISSN={1941-0042},
  month={},}

@InProceedings{Papadopoulos_2022_CVPR,
    author    = {Papadopoulos, Dim P. and Mora, Enrique and Chepurko, Nadiia and Huang, Kuan Wei and Ofli, Ferda and Torralba, Antonio},
    title     = {Learning Program Representations for Food Images and Cooking Recipes},
    booktitle = {Proceedings of the IEEE/CVF Conference on Computer Vision and Pattern Recognition (CVPR)},
    month     = {June},
    year      = {2022},
    pages     = {16559-16569}
}

@ARTICLE{9793670,
  author={Wang, Hao and Lin, Guosheng and Hoi, Steven C. H. and Miao, Chunyan},
  journal={IEEE Transactions on Pattern Analysis and Machine Intelligence}, 
  title={Learning Structural Representations for Recipe Generation and Food Retrieval}, 
  year={2023},
  volume={45},
  number={3},
  pages={3363-3377},
  doi={10.1109/TPAMI.2022.3181294},
  ISSN={1939-3539},
  month={March},}

@ARTICLE{9594708,
  author={Zhu, Bin and Ngo, Chong-Wah and Chan, Wing-Kwong},
  journal={IEEE Transactions on Multimedia}, 
  title={Learning From Web Recipe-Image Pairs for Food Recognition: Problem, Baselines and Performance}, 
  year={2022},
  volume={24},
  number={},
  pages={1175-1185},
  doi={10.1109/TMM.2021.3123474},
  ISSN={1941-0077},
  month={},}

@INPROCEEDINGS{10555071,
  author={Pitsilou, Vasiliki and Papadakis, George and Skoutas, Dimitrios},
  booktitle={2024 IEEE 40th International Conference on Data Engineering Workshops (ICDEW)}, 
  title={Using LLMs to Extract Food Entities from Cooking Recipes}, 
  year={2024},
  volume={},
  number={},
  pages={21-28},
  doi={10.1109/ICDEW61823.2024.00008},
  ISSN={2473-3490},
  month={May},}

@inproceedings{10.1145/3607828.3617797,
author = {Rodr\'{\i}guez-de-Vera, Jes\'{u}s M. and Villacorta, Pablo and Estepa, Imanol G. and Bola\~{n}os, Marc and Saras\'{u}a, Ignacio and Nagarajan, Bhalaji and Radeva, Petia},
title = {Dining on Details: LLM-Guided Expert Networks for Fine-Grained Food Recognition},
year = {2023},
isbn = {9798400702846},
publisher = {Association for Computing Machinery},
address = {New York, NY, USA},
url = {https://doi.org/10.1145/3607828.3617797},
doi = {10.1145/3607828.3617797},
booktitle = {Proceedings of the 8th International Workshop on Multimedia Assisted Dietary Management},
pages = {43–52},
numpages = {10},
location = {Ottawa ON, Canada},
series = {MADiMa '23}
}

@inproceedings{ming2018food,
  title={Food photo recognition for dietary tracking: System and experiment},
  author={Ming, Zhao-Yan and Chen, Jingjing and Cao, Yu and Forde, Ciar{\'a}n and Ngo, Chong-Wah and Chua, Tat Seng},
  booktitle={MultiMedia Modeling: 24th International Conference, MMM 2018, Bangkok, Thailand, February 5-7, 2018, Proceedings, Part II 24},
  pages={129--141},
  year={2018},
  organization={Springer}
}

@article{gao2019hierarchical,
  title={Hierarchical attention network for visually-aware food recommendation},
  author={Gao, Xiaoyan and Feng, Fuli and He, Xiangnan and Huang, Heyan and Guan, Xinyu and Feng, Chong and Ming, Zhaoyan and Chua, Tat-Seng},
  journal={IEEE Transactions on Multimedia},
  volume={22},
  number={6},
  pages={1647--1659},
  year={2019},
  publisher={IEEE}
}

@ARTICLE{8998172,
  author={Jiang, Landu and Qiu, Bojia and Liu, Xue and Huang, Chenxi and Lin, Kunhui},
  journal={IEEE Access}, 
  title={DeepFood: Food Image Analysis and Dietary Assessment via Deep Model}, 
  year={2020},
  volume={8},
  number={},
  pages={47477-47489},
  doi={10.1109/ACCESS.2020.2973625}}

@ARTICLE{10019590,
  author={Min, Weiqing and Wang, Zhiling and Liu, Yuxin and Luo, Mengjiang and Kang, Liping and Wei, Xiaoming and Wei, Xiaolin and Jiang, Shuqiang},
  journal={IEEE Transactions on Pattern Analysis and Machine Intelligence}, 
  title={Large Scale Visual Food Recognition}, 
  year={2023},
  volume={45},
  number={8},
  pages={9932-9949},
  doi={10.1109/TPAMI.2023.3237871}}

@ARTICLE{9846887,
  author={Wang, Zhiling and Min, Weiqing and Li, Zhuo and Kang, Liping and Wei, Xiaoming and Wei, Xiaolin and Jiang, Shuqiang},
  journal={IEEE Transactions on Image Processing}, 
  title={Ingredient-Guided Region Discovery and Relationship Modeling for Food Category-Ingredient Prediction}, 
  year={2022},
  volume={31},
  number={},
  pages={5214-5226},
  doi={10.1109/TIP.2022.3193763}}

@ARTICLE{9082900,
  author={Lo, Frank Po Wen and Sun, Yingnan and Qiu, Jianing and Lo, Benny},
  journal={IEEE Journal of Biomedical and Health Informatics}, 
  title={Image-Based Food Classification and Volume Estimation for Dietary Assessment: A Review}, 
  year={2020},
  volume={24},
  number={7},
  pages={1926-1939},
  doi={10.1109/JBHI.2020.2987943}}

@ARTICLE{10494044,
  author={Huang, Xu and Liu, Jin and Zhang, Zhizhong and Xie, Yuan and Tang, Yongqiang and Zhang, Wensheng and Cui, Xiaohui},
  journal={IEEE Transactions on Multimedia}, 
  title={Cross-Modal Recipe Retrieval With Fine-Grained Prompting Alignment and Evidential Semantic Consistency}, 
  year={2025},
  volume={27},
  number={},
  pages={2783-2794},
  doi={10.1109/TMM.2024.3384672}}

@ARTICLE{10471331,
  author={Liu, Yuxin and Min, Weiqing and Jiang, Shuqiang and Rui, Yong},
  journal={IEEE Transactions on Image Processing}, 
  title={Convolution-Enhanced Bi-Branch Adaptive Transformer With Cross-Task Interaction for Food Category and Ingredient Recognition}, 
  year={2024},
  volume={33},
  number={},
  pages={2572-2586},
  doi={10.1109/TIP.2024.3374211}}

@InProceedings{Wu_2024_CVPR,
    author    = {Wu, Xiongwei and Yu, Sicheng and Lim, Ee-Peng and Ngo, Chong-Wah},
    title     = {OVFoodSeg: Elevating Open-Vocabulary Food Image Segmentation via Image-Informed Textual Representation},
    booktitle = {Proceedings of the IEEE/CVF Conference on Computer Vision and Pattern Recognition (CVPR)},
    month     = {June},
    year      = {2024},
    pages     = {4144-4153}
}

\end{document}